\documentclass[fleqn,usenatbib]{mnras}

\usepackage{newtxtext,newtxmath}
\usepackage[T1]{fontenc}

\DeclareRobustCommand{\VAN}[3]{#2}
\let\VANthebibliography\thebibliography
\def\thebibliography{\DeclareRobustCommand{\VAN}[3]{##3}\VANthebibliography}

\usepackage{graphicx}	% Including figure files
\usepackage{amsmath}	% Advanced maths commands
\newcommand{\oiii}{[\ion{O}{iii}]}
\newcommand{\nii}{[\ion{N}{ii}]}
\newcommand{\sii}{[\ion{S}{ii}]}
\title[NFM observations of Circinus]{Dissecting the active galactic nucleus in Circinus IV. MUSE NFM observations unveil a tuning-fork ionised outflow morphology}

\author[D. Kakkad et al.]{
D. Kakkad$^{1}$,\thanks{E-mail: dkakkad@stsci.edu}
M. Stalevski$^{2,3}$,
M. Kishimoto$^{4}$,
S. Kne\v{z}evi\'{c}$^{2}$,
D. Asmus$^{5,6}$,
F.~P.~A. Vogt$^{7}$
\\
$^{1}$Space Telescope Science Institute, 3700 San Martin Drive, Baltimore, MD 21218, USA\\
$^{2}$Astronomical Observatory, Volgina 7, 11060 Belgrade, Serbia\\
$^{3}$Sterrenkundig Observatorium, Universiteit Gent, Krijgslaan 281-S9, Gent, 9000, Belgium\\
$^{4}$Department of Astrophysics \& Atmospheric Sciences, Faculty of Science, Kyoto Sangyo University, Kamigamo-motoyama, Kita-ku, Kyoto 603-8555, Japan\\
$^{5}$Department of Physics \& Astronomy, University of Southampton, Southampton, SO17 1BJ, UK\\
$^{6}$Gymnasium Schwarzenbek, 21493 Schwarzenbek, Germany\\
$^{7}$Federal Office of Meteorology and Climatology - MeteoSwiss, Chemin de l'Aérologie 1, 1530 Payerne, Switzerland\\
}

\date{Accepted XXX. Received YYY; in original form ZZZ}

\pubyear{2021}

\begin{document}
\label{firstpage}
\pagerange{\pageref{firstpage}--\pageref{lastpage}}
\maketitle

% Abstract of the paper
\begin{abstract}
We present the ionised gas outflow morphology in the Circinus galaxy using the Narrow Field Mode (NFM) of the MUSE instrument on board the Very Large Telescope (VLT). The NFM observations provide a spatial resolution of $\sim$0.1\arcsec, corresponding to a physical scale of $\sim$2 pc, one of the highest spatial resolution achievable using ground-based AO-assisted observations in the optical wavelengths. The MUSE observations reveal a collimated clumpy outflow profile originating near the AGN location and extending up to 1.5\arcsec ~($\sim$30 pc) in the NW direction. The collimated structure then fragments into two filaments, giving the entire outflowing gas a ``tuning-fork'' morphology. These structures remain undetected in the lower spatial resolution MUSE Wide Field Mode data. We explain the origin of this tuning-fork structure to the interaction of the outflow with a dense clump in the interstellar medium (ISM) as the outflow propagates outward. The origin of the collimated structure itself could be from jet-ISM interactions on small scales. These observations also provide evidence to the origin of the ionised gas filaments previously observed in the Circinus galaxy out to kiloparsec scales. We find instantaneous and time-averaged mass outflow rates of 10$^{-2}$ M$_{\odot}$ yr$^{-1}$ and 10$^{-4}$ M$_{\odot}$ yr$^{-1}$, respectively. Based on the star formation rate in the Circinus galaxy reported in the literature, the observed ionised outflows are not expected to regulate star formation within the $\sim$100 pc scales probed by the NFM data.
\end{abstract}

% Select between one and six entries from the list of approved keywords.
% Don't make up new ones.

%%%%%%%%%%%%%%%%%%%%%%%%%%%%%%%%%%%%%%%%%%%%%%%%%%

%%%%%%%%%%%%%%%%% BODY OF PAPER %%%%%%%%%%%%%%%%%%

\section{Introduction} \label{sect1}

The so-called Unified Model (UM) of the active galactic nuclei (AGN) consists of a central black hole surrounded by an equatorial torus-like structure, which is responsible for the angle-dependent obscuration of the accretion disk and in some cases, may include collimated jets along the polar directions \citep[e.g.,][]{antonucci93, urry95, netzer15}. The equatorial torus has been believed to dominate the infrared emission from the AGN \citep[see][and the references therein]{ramos-almeida17}. There have been intensive efforts in the literature, both from an observational as well as modelling perspective, to study the nature of this torus, such as its geometry \citep[e.g.,][]{honig19,garcia-burillo21}, what are the typical dust covering factors \citep[e.g.,][]{elitzur12, stalevski16, toba21} and whether the material is clumpy or smooth \citep[e.g.,][]{dullemond05, marin15, garcia-gonzalez17}. 

High resolution mid-infrared observations over the past decade have now challenged these simplified torus model that have dusty clumps only in the equatorial region \citep{nenkova02}. Several studies in the literature now show strong infrared emission along the polar direction on the scales of a few parsecs \citep[e.g.,][]{honig12, honig13, lopez-gonzaga16, honig17, leftley18} to hundreds of parsecs \citep[e.g.,][]{braatz93, bock00, asmus16, asmus19}. As a result, the dust emission around the AGN is believed to consist of two components: an equatorial thin disk and a polar extended feature that could originate from the winds from the central engine \citep[see][and the references therein]{honig19}. 

The polar wind is believed to have a multi-phase composition ranging from dust to ionised and molecular gas components. In fact, recent high spatial resolution observations of nearby AGN have targeted the molecular gas distribution around the torus using ALMA, revealing high velocity outflows in the molecular gas phase \citep[e.g.,][]{gallimore16, combes19, garcia-burillo19, lopez-rodriguez20}. In order to get a holistic view of the multi-phase dusty gas flows around the torus, it is imperative to obtain the morphology and kinematics of the ionised gas on the same scales as the molecular gas and infrared emission. Furthermore, obtaining outflow morphology at such small spatial scales can also give clues into the outflow launching mechanism and connect them to the observed kiloparsec scales structure or outflows, whenever available in the literature. Thanks to the Narrow Field Mode (NFM) capabilities of the Multi Unit Spectroscopic Explorer \citep[MUSE][]{bacon10} at the Very Large Telescope (VLT), such high spatial resolution observations can now be performed with ground-based Integral Field Spectroscopic instruments operating at optical wavelengths. The optical wavelengths provide access to bright emission lines such as the \oiii$\lambda$5007 that trace ionised gas in the Narrow Line Region (NLR). Furthermore, emission lines such as H$\alpha$, H$\beta$, \nii$\lambda\lambda$6549, 6585 and \sii$\lambda\lambda$6716, 6731 help derive dust extinction maps and diagnostic diagrams that trace the source of ionisation across the field-of-view. 

\begin{figure*}
\centering
\includegraphics[scale=0.5]{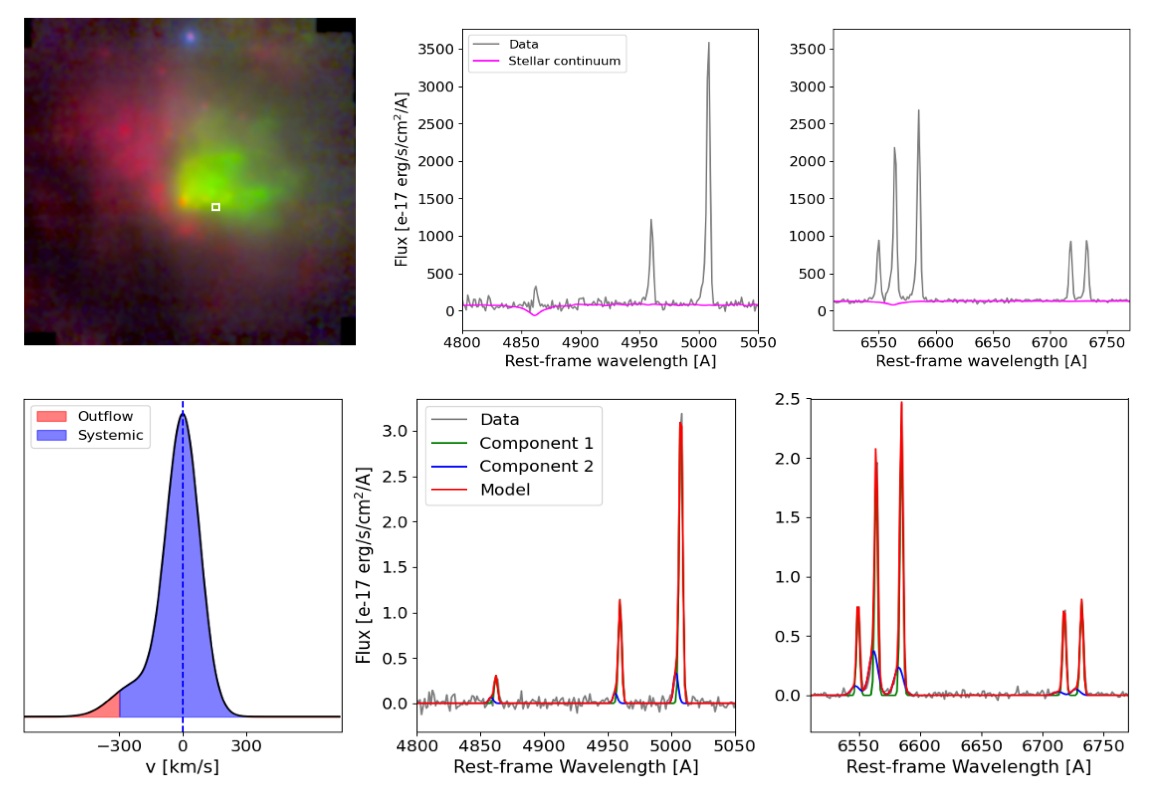}
\caption{The top left panel shows an RGB colour image of the Circinus galaxy, derived from the NFM data. The field-of-view of the RGB image is 7.5\arcsec$\times$7.5\arcsec. The red hue indicated the H$\alpha$ emission, while the ionised gas cone is apparent in green. The white square at the edge of the ionisation cone shows the pixel from where the example spectra shown in this figure was obtained. The middle and right panels in the top row shows the example of a stellar continuum fit and the middle and right panels in the bottom row shows the emission line fitting results of H$\beta$, \oiii$\lambda\lambda$4959,5007, \nii$\lambda\lambda$6549,6585, H$\alpha$ and \sii$\lambda\lambda$6716,6731. In all the spectra, the grey curve shows the extracted data from the pixel location shown in the median image, the magenta curve shows the stellar continuum model, the green and blue curves show the individual Gaussian components used to model the emission lines and the red curve shows the overall model. The bottom left panel illustrates the definition of outflow and systemic flux used in this paper. Further details are given in Sect. \ref{sect3}.}
\label{fig:illustration}
\end{figure*}

The Circinus galaxy is the closest Seyfert 2 galaxy \citep[$\sim$4.2 Mpc away, z = 0.001][]{freeman77} and hosts an infrared-bright AGN \citep[e.g.,][]{jarrett19}. The polar axis of the AGN in the Circinus galaxy is seen almost edge-on and is therefore an ideal target to study the relative gas and dust structure in a typical obscured AGN. Recent mid-infrared (MIR) observations using the upgraded VISIR instrument at the Very Large Telescope (VLT) suggests the presence of dust in the form of a hollow cone at the edges of the ionised outflow \citep[e.g.,][]{stalevski17}. The polar elongation of the infrared emission has also been reported on parsec scales \citep[e.g.,][]{tristram14, stalevski19} using Mid-Infrared Interferometric Instrument (MIDI) and the Multi AperTure mid-infrared Spectro-Scopic Experiment \citep[MATISSE, e.g.,][]{isbell22} at the VLT. The presence of dust in the form of a hollow cone in the polar region is also confirmed by multi-band optical polarimetry with VLT/FORS2 (Stalevski et al., submitted) The galaxy also hosts powerful outflows in the ionised gas phase, visible in the form of a one-sided ionisation cone extended up to $\sim$1 kiloparsec \citep[e.g.,][]{marconi94, veilleux97, mingozzi19, fonseca-faria21, kakkad22}. The Circinus galaxy hosts an obscured AGN with nuclear star formation that dominates the dust emission on scales of hundreds of parsec \citep[e.g.,][]{matt00, arevalo14}.

In this fourth paper in the series, we map the morphology and kinematics of ionised gas in the Circinus galaxy at $\sim$2 pc resolution using MUSE-NFM observations. We present a model of the ionisation cone and the resulting outflowing gas structure. The observed ionised outflow morphology obtained from the NFM observations is compared with the larger scale outflows observed with the Wide Field Mode (WFM) of MUSE, to understand outflow propagation across the host galaxy. We locate the regions with high dust extinction and compare this with the archival mid-infrared images. Lastly, the MUSE data is compared with other archival radio observations to infer the presence of jet-ISM interaction in the host galaxy. 

Throughout this paper, e adopt the following $\Lambda$CDM cosmological parameters: $H_{0} = 70$ km s$^{-1}$, $\Omega_{\rm M} = 0.3$ and $\Omega_{\Lambda} = 0.7$. All the maps use the following convention: North is up and East is to left. 

\section{MUSE-NFM observations \& data reduction} \label{sect2}

The observations were performed using the Laser Tomographic Adaptive Optics (LTAO) assisted Narrow Field Mode of the MUSE instrument, on board Unit Telescope 4 of the VLT\footnote{ESO programme ID: 0103.B-0396(A)}. The observations were carried out on the nights of 29 and 30 April 2019 with a DIMM seeing in the range 0.89--1.01\arcsec. We observed the galaxy with an optimised sequence: O-S-O-O-S-O (O = Object, S = Sky), where the Sky was obtained at an offset position $\sim$1.5 arcmin away outside the galaxy. We performed small dithering between the individual science exposures and rotated the field by 90 degrees on each subsequent exposure to eliminate the impact of bad pixels and to average out the patterns of slicers and channels. The nucleus of the Circinus galaxy has an H-band magnitude of 13.4 and therefore, served as the Adaptive Optics (AO) reference for the Wavefront Sensor (WFS). The total on-source exposure time was $\sim$4000s.

The raw data was reduced using the standard MUSE pipeline \citep[e.g.,][]{weilbacher14, weilbacher20}. The pipeline performs bias correction, flat fielding, wavelength and astrometry calibration, sky subtraction and flux calibration. The final data cube consisted of a field-of-view of $\sim$7.5$\times$7.5 arcsec$^{2}$ centred on the nucleus (AGN) with a spatial sampling of 0.025\arcsec. As the observations were performed in the Nominal mode, this provided a uniform wavelength coverage between 4800--9300 \r{A} with a gap between 5780--6050 \r{A} due to the presence of a notch filter that suppresses the Sodium laser light. The spectral PSF (also known as the Line Spread Function, LSF) is in the range 2.5--2.9 \r{A}, with the best resolution obtained at the redder end of the spectra. This corresponds to a velocity resolution of $\sim$150 km s$^{-1}$ at the location of \oiii$\lambda$5007 line, one of the emission lines that will be analysed in this paper. The LTAO-assisted observations resulted in a spatial PSF of $\sim$0.1\arcsec, determined using one of the point sources in the field-of-view. This spatial resolution is one of the highest that can be achieved using ground-based IFS observations. At the redshift of the Circinus, this corresponds to a physical scale of $\sim$2 pc, which means that the observations can potentially resolve the region near the AGN torus. With a field-of-view of 7.5\arcsec$\times$7.5\arcsec, the NFM observations trace spatial scales up to $\sim$100 pc from the AGN location.

\section{Analysis} \label{sect3}
In order to derive the flux and velocity maps from the optical emission lines, we first subtract the stellar continuum emission across the MUSE field-of-view. We perform the stellar continuum emission using the LZIFU tool \citep[e.g.,][]{ho16, kreckel18}, which adopts the penalized pixel fitting routine \citep[PPXF][]{cappellari04, cappellari17} to fit the stellar continuum using input stellar spectrum templates (from \citet{gonzalez-delgado05}) or modelled simple stellar populations (SSPs) that are convolved with parametrised velocity distributions. In doing so, regions in the spectra with strong skylines and emission lines intrinsic to the host galaxy were masked. The key emission lines that were masked were H$\beta$, \oiii$\lambda\lambda$4959, 5007, \nii$\lambda\lambda$6549, 6585, H$\alpha$ and \sii$\lambda\lambda6716, 6731$ (\sii ~doublet hereafter). We also mask the notch filter region in the spectrum that is contaminated by the sodium doublet emission from the lasers. Being a Seyfert 2 galaxy, the Circinus galaxy does not display broad emission lines from the Broad Line Region (BLR). 

The resulting stellar continuum-subtracted data cubes were then used to analyse the morphology, kinematics and the ionisation mechanism of the gas using strong emission lines in the optical spectra. For instance, we used the \oiii$\lambda$5007 line (\oiii ~hereafter) to trace the ionised gas in the Narrow Line Region (NLR), the Balmer lines H$\alpha$ and H$\beta$ lines are used to trace potential star formation and dust extinction in the host galaxy (using Balmer decrement), the \nii$\lambda\lambda$6549, 6585 lines are used in investigating the ionisation source in each pixel (AGN or star formation) and the \sii ~doublet are used to trace the ionised gas electron density. We model these emission lines with multiple Gaussian functions using the \texttt{scipy.curve-fit} package in \texttt{python} \citep[see][]{virtanen20}. Initially a single Gaussian is fitted to the emission line profile, and additional Gaussian functions were added if the $\chi^{2}$ is minimised, and until the line fluxes are stable within $\sim$10\%. Circinus does not display BLR emission in its H$\beta$ or H$\alpha$ profiles and the maximum number of Gaussians required to model an emission line was two. We do not assign any physical significance to the individual Gaussian components as we follow a non-parametric approach to derive the properties associated with the systemic and outflow components. The non-parametric approach was chosen over a parametric model as it does not depend on the choice of the models used for the emission lines (e.g., Gaussian, Lorentzian or a power-law). In addition, we tied the line centroids of \oiii ~line with that of H$\beta$ and the \nii ~and \sii ~lines with that of H$\alpha$ based on the expected location of the respective atomic species. The emission line ratios \oiii$\lambda$5007:\oiii$\lambda$4959 and \nii$\lambda$6585:\nii$\lambda$6549 were set approximately equal to 3:1 based on expectations from theory \citep[e.g.,][]{osterbrock06, dimitrijevic07}. Lastly, the FWHM of the individual Gaussian components of the \oiii ~line was coupled with H$\beta$ and the H$\alpha$ FWHM with that of \nii. 

From the line fitting procedure, we are interested in the following parameters: The 10th percentile velocity: $v_{10}$ (blue-shifted \oiii ~velocity that contains 10\% of the overall \oiii ~flux), the width of the emission line: $w_{80}$ \citep[width containing 80\% of the flux of the emission line, see][]{harrison14, kakkad16,wylezalek20} and the flux of the outflowing and systemic components of each of the emission lines. To determine the fluxes, we first define the zero velocity location in the emission line spectra, which is the location of the peak of the emission line. The flux within $\pm$300 km s$^{-1}$ on either side of this zero velocity location is considered to be the systemic component of the emission line. The choice of 300 km s$^{-1}$ is based on the line width (FWHM) cut of $\sim$600 km s$^{-1}$ which is often employed in the literature to distinguish between outflowing and non-outflowing gas \citep[e.g.,][]{kakkad20}. Using similar arguments, we use the flux outside of the $\pm$300 km s$^{-1}$ channels to define the flux associated with outflows. Using lower velocity cuts such as 250 or 200 km s$^{-1}$ yield similar results, but due to the possibility of contamination from the non-outflowing gas at the lower velocities, we make a conservative cut of 300 km s$^{-1}$. Figure \ref{fig:illustration} shows an example of the analysis methods and the fitting results presented in this section. 

\begin{figure*}
\centering
\includegraphics[scale=0.4]{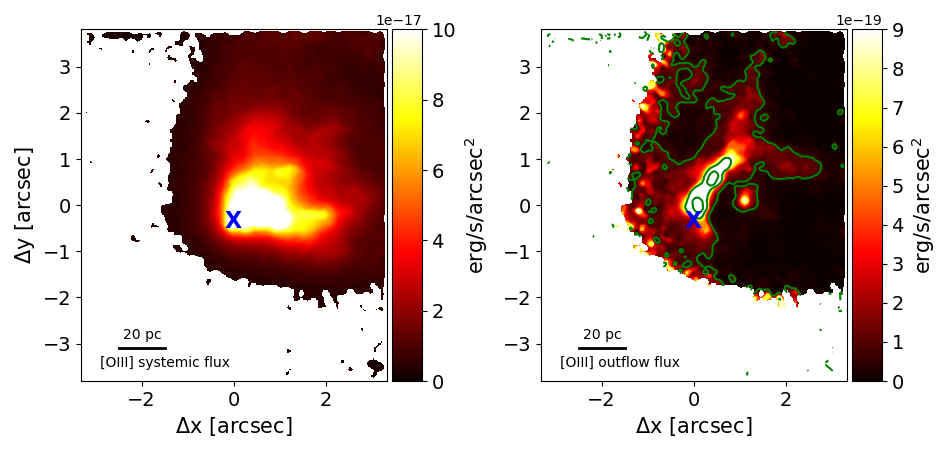}
\includegraphics[scale=0.38]{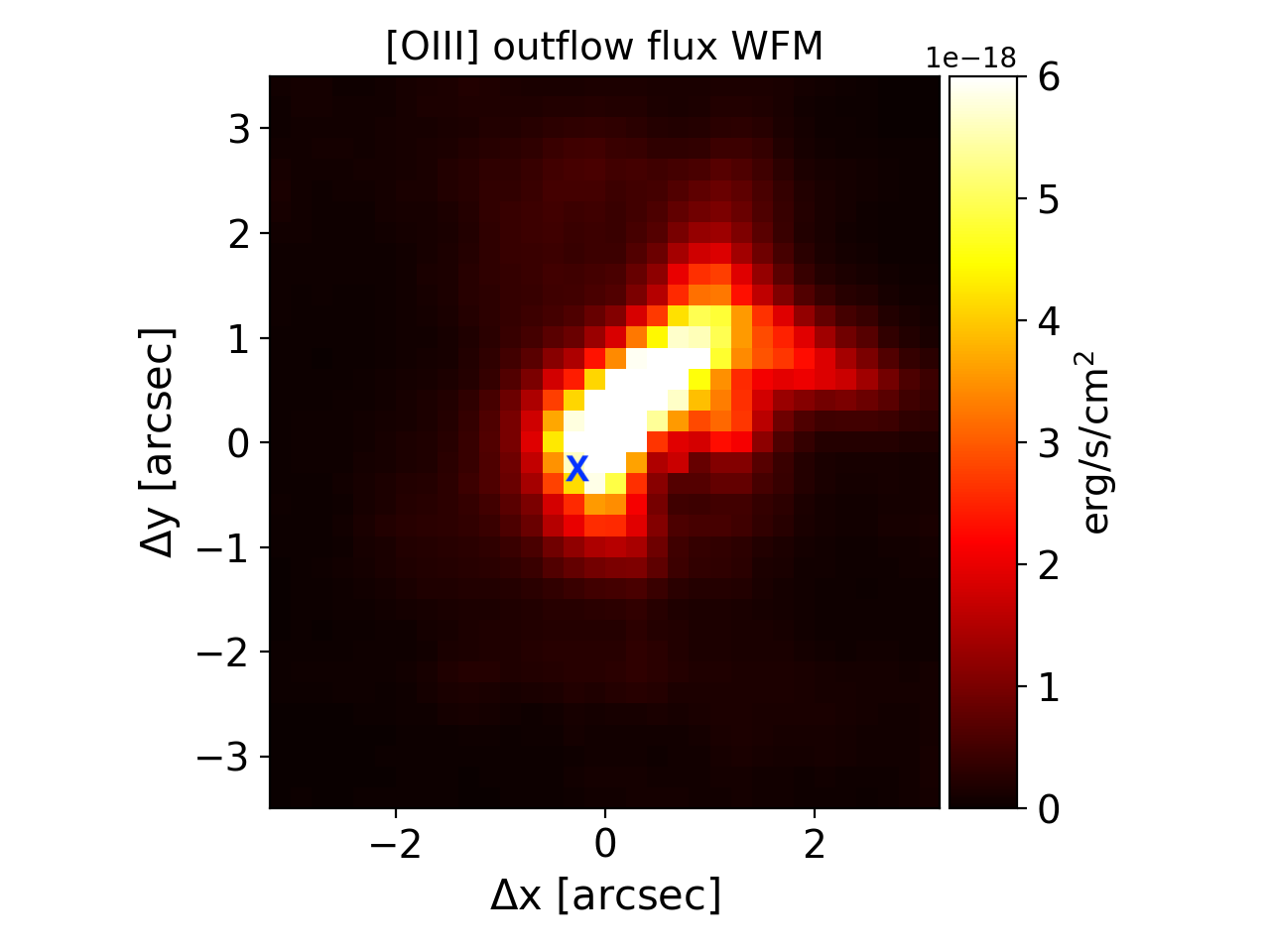}
\caption{The left panel shows the systemic \oiii ~flux ($\lvert v \rvert <$ 300 km s$^{-1}$) that traces the ionisation cone. The middle panel shows the outflowing component of the \oiii ~flux ($\lvert v \rvert >$ 300 km s$^{-1}$). The outflow morphology suggests the gas propagation along a collimated structure before it fragments into two filaments, giving it a "tuning-fork" resemblance. The green contours in the middle panel shows the \oiii ~outflow contours within the collimated structure to highlight that the outflowing structure itself shows the presence of clumps. Thanks to the 0.1\arcsec ~spatial resolution achieved with the NFM observations, such structures are barely visible in archival WFM data shown in the right panel. In all the panels, the blue "X" marks the AGN location and the black bar on the bottom left shows the 20 pc scale.}
\label{fig:outflow_flux}
\end{figure*}

\begin{figure*}
\centering
\includegraphics[scale=0.35]{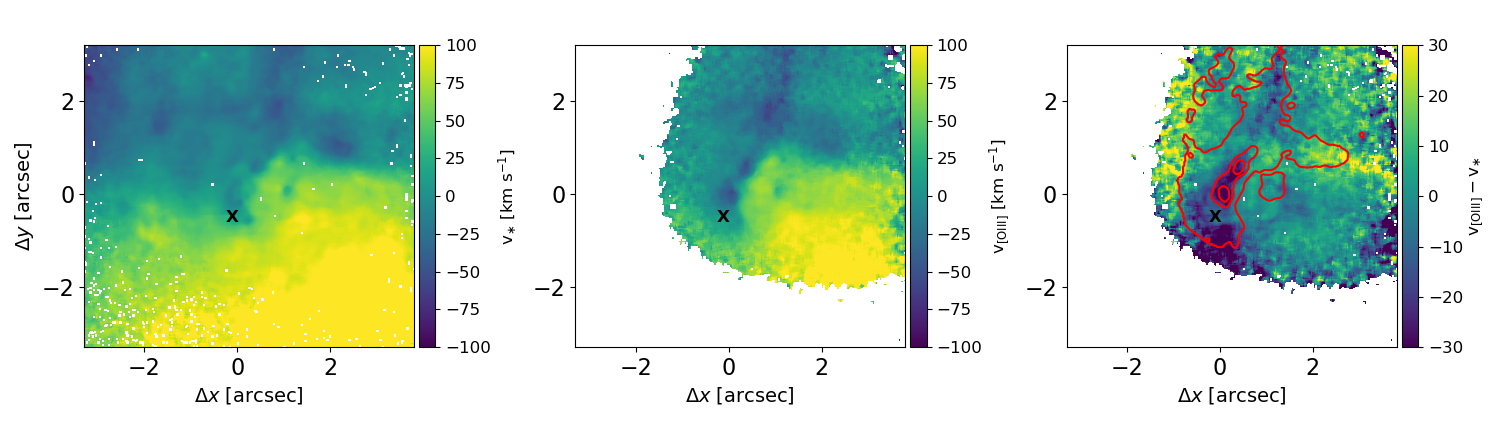}\\
\includegraphics[scale=0.35]{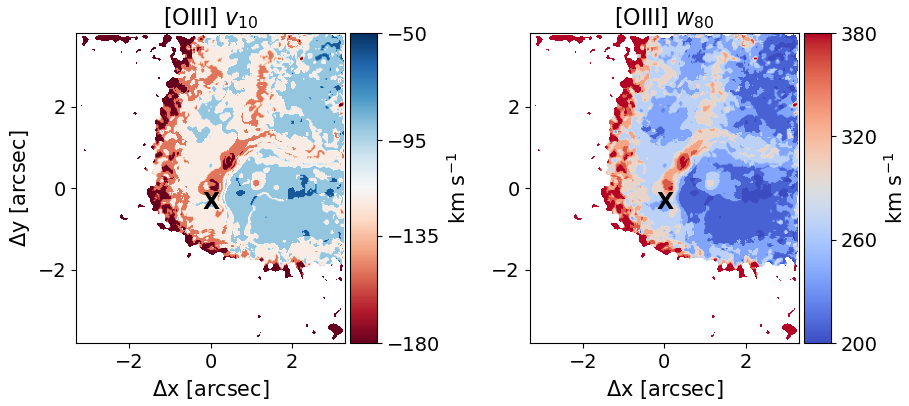}
\caption{The top left panel shows the stellar velocity map obtained from the stellar continuum fitting. The stellar velocity map shows a smooth rotation-like profile of the host galaxy. The \oiii ~centroid velocity profile (top centre panel) approximately mimics the stellar velocity field, suggesting that the bulk of the ionised gas cone co-rotates with the host galaxy. The top right panel shows the residual map after subtracting the stellar velocity map from the \oiii ~velocity map. We observe residuals at the locations of the "tuning-fork" structure (red contour), suggesting that it is a part of the non-rotation component. The positive residuals in the filament directed towards the West and the negative residuals in the filament towards North shows that the outflow itself is co-rotating with the ionised gas and the host galaxy. The bottom panels show the non-parametric velocities, $v_{10}$ and $w_{80}$, described in Sect. \ref{sect3}. Both these velocities confirm that the high velocity regions are along the collimated structure that fragments into two filaments $\sim$1.5\arcsec ~from the AGN location. Furthermore, the presence of this structure in the $v_{10}$ map shows that the dominant component of the outflow is blue-shifted. The black "X" in all maps indicate the AGN location.}
\label{fig:velocity_maps}
\end{figure*}

\section{Results \& discussion} \label{sect4}

In this section, we show the results of the analysis methods described in Sect. \ref{sect3}. The main aim of this section is to derive the ionised gas outflow morphology and kinematics close to the AGN torus and compare this with archival multi-wavelength data, including the low spatial resolution MUSE-WFM data.

\subsection{The parsec-scale ionised outflow in the Circinus galaxy} \label{sect4.1}

Figure \ref{fig:outflow_flux} shows the flux maps of the systemic (left panel) and outflowing component (middle panel) of the \oiii ~emission line from the NFM data. The systemic \oiii ~flux map traces the ionisation cone originating from the AGN location and extends towards the NW direction from the nucleus. The presence of the ionisation cone in the Circinus galaxy has previously also been reported in the literature, including MUSE Wide Field Mode (WFM) observations \citep[e.g.,][]{marconi94, fischer13, mingozzi19, fonseca-faria21, kakkad22}. The systemic flux dominates the bulk of the ionised gas flux in the host galaxy by nearly two orders of magnitude, compared to the \oiii ~outflow flux. The \oiii ~outflow map (middle panel, Fig. \ref{fig:outflow_flux}), on the other hand, shows a collimated structure that originates close to the AGN location and extends towards the NW of the nucleus (same direction and approximately the same PA as the ionisation cone). The collimated structure itself is not uniform and shows multiple clumps. Such clumps have also been previously reported in extended radial ionised gas filaments of the Circinus galaxy \citep[e.g.,][]{veilleux97}. We note that the location of the first clump is not coincident with the AGN location, but $\approx$0.4\arcsec ~NW of the nucleus. In Section \ref{sect5}, we further discuss the origin of these clumps and whether they could be produced by the shocks within the outflowing wind. 

Beyond $\sim$1.5\arcsec ~from the AGN location ($\sim$30 pc) in the NW direction, the collimated structure then fragments into two filaments, one towards the West and another towards North, which gives the overall outflow morphology a "tuning-fork" resemblance. The impact of the high resolution NFM observations is clear from these observations as such pc-scale filaments and fragmenting structures are not visible in the archival low resolution ($\sim$0.5\arcsec) MUSE WFM data, as shown in the right panel of Fig. \ref{fig:outflow_flux}. 

Figure \ref{fig:velocity_maps} shows velocity maps of the stellar component and the ionised gas of the Circinus galaxy, derived from the NFM observations. The top left panel in Fig. \ref{fig:velocity_maps} shows the stellar velocity distribution in the host galaxy obtained from the stellar continuum modelling. The velocity map shows a smooth gradient indicating a rotation-like profile, with the axis of rotation aligned approximately along the axis of the ionisation cone. The \oiii ~centroid map, shown in the top centre panel of Fig. \ref{fig:velocity_maps}, mimics the stellar velocity map i.e., the ionised gas co-rotates with the host galaxy. The \oiii ~centroid velocity profile is also consistent with previous MUSE-WFM results from the literature \citep[e.g.,][]{fonseca-faria21}. On subtracting the stellar velocity component from the \oiii ~centroid velocity, we see clear residuals at the locations of the "tuning-fork" structure, as shown in the top right panel of Fig. \ref{fig:velocity_maps}. This proves that the outflow flux shown in the middle panel in Fig. \ref{fig:outflow_flux} is indeed part of the non-systemic component of the host galaxy. Furthermore, the positive and negative residuals in the West and North arms respectively in the residual map in the top right panel of Fig. \ref{fig:velocity_maps} indicates that the fork structure itself is co-rotating with the host galaxy and the ionisation cone. The bottom panels in Fig. \ref{fig:velocity_maps} show the \oiii ~$v_{10}$ and the $w_{80}$ maps (left and right panels respectively). Both the $v_{10}$ and $w_{80}$ maps confirm the results seen in the outflow maps i.e., the high velocity regions are collimated up to $\sim$1.5\arcsec ~from the AGN location before they fragment into two filaments. The presence of the tuning-fork structure in the $v_{10}$ map suggests that most of the observed outflow flux is dominated by the blue-shifted emission. We note that the stellar velocity map in the top left panel of Fig. \ref{fig:velocity_maps} also shows a "V-shaped" structure at approximately the same location where the high velocity regions fragment into the two arms, suggesting that the material within the cone is both outflowing and co-rotating with the host galaxy. 

\begin{figure}
\centering
\includegraphics[scale=0.45]{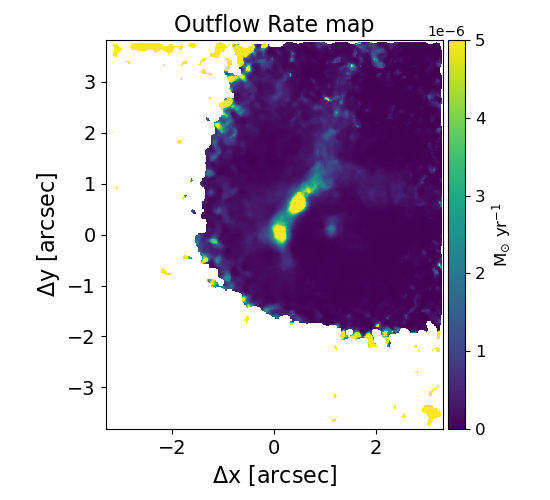}
\caption{The map shows the mass outflow rate distribution for the ionised gas derived from the \oiii$\lambda$5007 line in the MUSE-NFM observations of the Circinus galaxy. The mass outflow rates are higher in regions with higher outflow velocity.}
\label{fig:outflow_rate}
\end{figure}

We also derived the ionised gas mass outflow rate using the \oiii ~line adopting methods from the literature \citep[e.g.,][]{rupke05, genzel11, veilleux20, kakkad22}. We report two kinds of outflow rate values: Instantaneous outflow rates ($\dot{M}_{\rm inst}$) is the sum of mass outflow rates calculated for each pixel, and time-averaged mass outflow rate ($M_{\rm Tavg}$) calculated by taking averaged quantities over the whole outflowing region. These quantities can be computed using the following equations:

\begin{equation}
\dot{M}_{\rm inst} = \sum_{\rm pix} \frac{M_{\rm out}\cdot v_{\rm out}}{\Delta R}
\label{eq:MOR_inst}
\end{equation}

\begin{equation}
\dot{M}_{\rm Tavg} = \frac{M_{\rm out}\cdot <v_{\rm out}>}{R}
\label{eq:MOR_Tavg}
\end{equation}

In Equation \ref{eq:MOR_inst}, the mass of the outflowing ionised gas, $M_{\rm out}$ and the velocity of the ionised gas, $v_{\rm out}$ is computed for each pixel and $\Delta R$ is the size of the pixel. In Equation \ref{eq:MOR_Tavg}, on the other hand, $M_{\rm out}$ is the total outflowing gas mass computed from the outflowing \oiii ~flux and $v_{\rm out}$ is the average velocity over the outflowing region ($\sim$300 km s$^{-1}$). The parameter, $R$, in Eq. \ref{eq:MOR_Tavg} is the distance of the outflow from the AGN location. As we are using spatially-resolved observations, we do not need to assume an outflow geometry or outflow density for the time-averaged quantity. The outflow density in both cases is obtained from the flux ratio of the outflowing components of \sii$\lambda\lambda$6716, 6731. 

The mass outflow rate map, representing the instantaneous mass outflow rates (Eq. \ref{eq:MOR_inst}), is shown in Fig. \ref{fig:outflow_rate}, where the mass outflow rate was calculated for each pixel. The advantage of using this method is that variation in the outflow parameters such as outflow density and velocity can be incorporated without the need for any assumptions on outflow models. We find the median outflow density across the field-of-view, calculated using the flux ratio of \sii$\lambda\lambda$6716, 6731 to be $\sim$200 cm$^{-3}$ \citep[e.g.,][]{sanders16, kaasinen17, kakkad18}. The total summed instantaneous outflow rate is 0.01 M$_{\odot}$ yr$^{-1}$ (an average of 3$\times$10$^{-7}$ M$_{\odot}$ yr$^{-1}$ per pixel where outflow is detected), which is two orders of magnitude less than the total instantaneous outflow rate value reported with MUSE-WFM observations \citep{kakkad22}. The time-averaged outflow rate computed using Eq. \ref{eq:MOR_Tavg} is 10$^{-4}$ M$_{\odot}$ yr$^{-1}$. The obscured star formation rate (SFR) in the Circinus galaxy is reported to be between 3--8 M$_{\odot}$ yr$^{-1}$. The orders of magnitude difference between the SFR and the ionised outflow rate within a radius of $\sim$100 pc of the AGN location, therefore, suggests that the observed ionised outflow is not expected to shut down star formation in the host galaxy. However, this may not be true for kiloparsec-scale molecular outflows where the outflow rate in the molecular gas phase has been reported to be $\sim$0.35--12.3 M$_{\odot}$ yr$^{-1}$ \citep[see][]{zschaechner16}. The high molecular outflow rate in kiloparsec-scales can, therefore, regulate star formation. A multi-phase approach to high resolution gas kinematics, by tracing warm and cold molecular gas components, is required to robustly confirm whether these AGN outflows affect star formation within $\sim$100 pc of the AGN. 

\begin{figure*}
\centering
\includegraphics[scale=0.4]{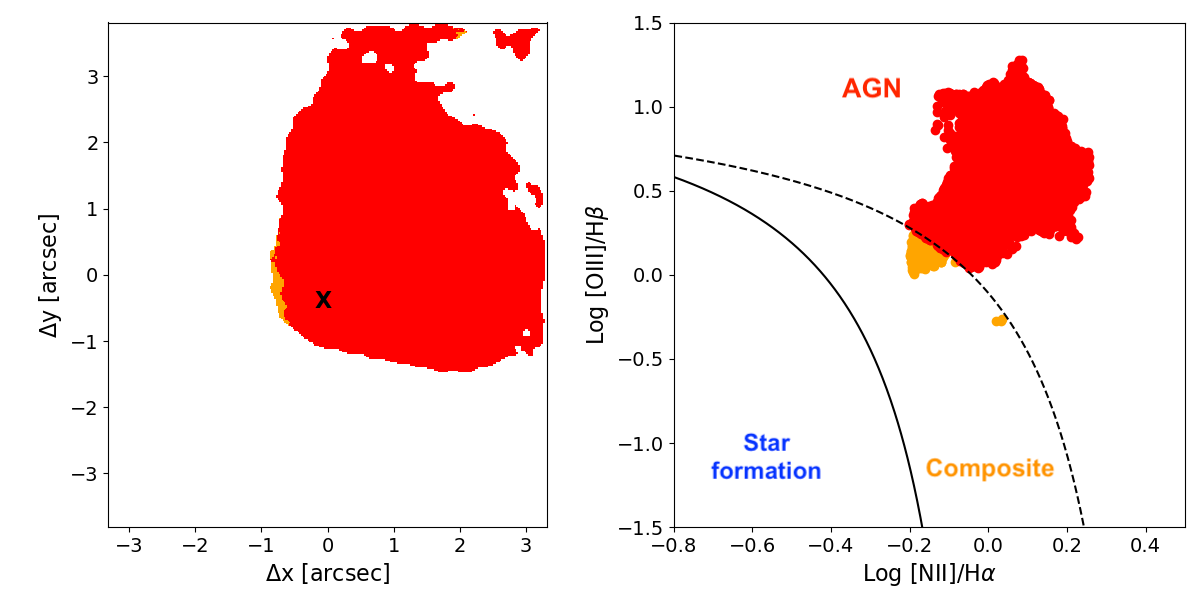}
\caption{The left panel shows the ionisation structure (AGN in red and composite in orange) in the field-of-view probed by the MUSE-NFM data. The right panel shows the location of each pixel in the classical \nii ~BPT diagram. The solid and dashed black curves are obtained from \citet{kauffmann03} and \citet{kewley01} and divide the plots between regions ionised by AGN, star formation and composite processes. The systemic flux of the emission lines was used while plotting this diagram. However, the results are similar if the outflowing components is used. The figure highlights that the gas is ionised primarily by the AGN.}
\label{fig:NFM_BPT}
\end{figure*}

Lastly, using spatially resolved Baldwin, Phillips \& Terlevich diagrams \citep[e.g.,][]{baldwin81, veilleux87}, we infer that the dominant source of ionisation across the NFM field-of-view is the AGN and the ionisation by star formation is negligible or absent (Figure \ref{fig:NFM_BPT}). The ionisation structure is consistent with previous WFM results in the literature \citep[e.g.,][]{mingozzi19, kakkad22}. The ionisation by the AGN is observed for both systemic as well as outflowing components. Therefore, the current observations also do not support a scenario where these outflows trigger star formation activity in the vicinity of the AGN. 

\subsection{The dust-outflow connection in Circinus} \label{sect4.2}

\begin{figure}
\centering
\includegraphics[scale=0.45]{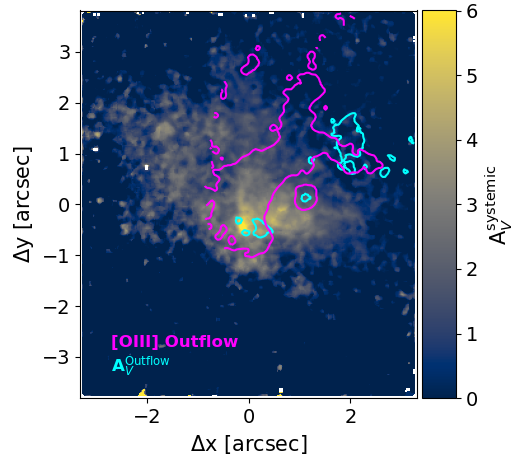}
\caption{Dust extinction ($A_{\rm V}$) map of the Circinus galaxy using the NFM observations. The background image shows the extinction map obtained from the systemic components of H$\alpha$ and H$\beta$, the cyan contours show the extinction from the outflowing components and the magenta contours show the location of high velocity ionised gas outflow. The dust extinction is dominant along the polar direction, consistent with previous mid-infrared observations in the literature.}
\label{fig:extinction}
\end{figure}

Previous mid-infrared observations of the Circinus galaxy established that a major fraction of dust emission is coming from the polar region, tentatively associated with dusty winds driven by radiation pressure \citep[e.g.,][]{stalevski17, venanzi20}. Even though far away from the central engine, dust and gas are expected to be coupled and co-spatial, and until recently the models of infrared emission ignored this polar dust component.  The spatially-resolved optical spectra from the MUSE-NFM mode can be used to derive extinction maps from Balmer decrement (H$\alpha$/H$\beta$) to confirm the presence of dust along the polar direction. Therefore, we derived the host galaxy extinction, $A_{\rm V}$, across the NFM field-of-view using the Balmer Decrement parameter. We assumed a \citet{calzetti00} dust attenuation law with $R_{\rm V}$ = 4.05 and a fixed temperature of 10,000 K, which is the typical electron temperature in the NLR. We note that the Circinus galaxy suffers from Galactic extinction of $A_{\rm V} \sim$2 \citep[see][]{for12}. However, the \oiii ~outflow morphology and the associated velocities and mass outflow rates will not change on correcting the Galactic extinction. The extinction map is shown in Fig. \ref{fig:extinction}. The background map in Fig. \ref{fig:extinction} shows the extinction map obtained from the systemic components of the flux ratio, H$\alpha$/H$\beta$. The cyan contours show the extinction ($A_{\rm V}>1$) obtained from the outflowing component of H$\alpha$ and H$\beta$, and the magenta contours show the location of the \oiii ~ionised gas outflow from the middle panel in Figure \ref{fig:outflow_flux}.

While the map in Fig. \ref{fig:extinction} shows potential dust distribution both along the disk (consistent with the results reported in \citet{mingozzi19} and \citet{fonseca-faria21}) as well as the polar direction, the overall distribution is dominant along the polar direction. This result is consistent with the previous results with mid-infrared emission, which was also dominant along the polar direction \citep[e.g.,][]{jaffe04, tristram07, tristram14, asmus16}. This suggests that the dust along the polar direction might be a part of lower velocity gas (compared to the high velocity collimated outflow observed here) surrounding the ionised gas outflow. The extinction from the systemic components peaks at the location of the AGN and gradually falls off to $A_{\rm V}$ = 0 at a distance of $\approx$3\arcsec ~i.e., $\approx$60 pc. 

The extinction obtained from the outflowing H$\alpha$ and H$\beta$ components shows non-uniform clumps scarcely distributed along the polar direction. We attribute these clumps to be a part of the outflowing gas and dust. It is worth noting that one of these clumps is almost at the tip of the collimated component of the outflow, approximately where the outflow filament fragments into two arms. The observation, therefore, might support a picture where the ionised gas outflow chooses the path of least resistance and therefore fragments into the two filaments, avoiding the region radially outward towards where the dust clump is present.

\section{Discussion} \label{sect5}

\begin{figure}
\centering
\includegraphics[scale=0.5]{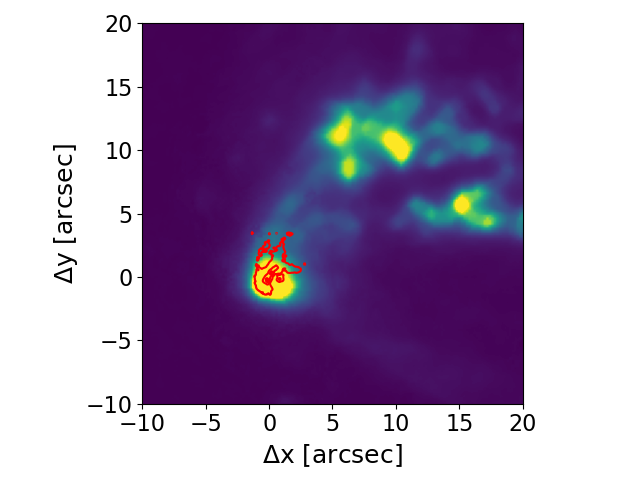}
\caption{The background image shows the \oiii$\lambda$5007 flux map from the MUSE-WFM data, which shows an overall asymmetric conical morphology with two distinct filaments on hundreds of parsec scales. The red contours show the \oiii ~outflow morphology obtained from the NFM observations in the middle panel of Fig. \ref{fig:outflow_flux}. Comparing the NFM contours with that of the WFM \oiii ~flux map, it appears that the two filaments observed in the WFM data have their origins in parsec-scale outflow traced with the NFM.}
\label{fig:NFM_WFM_connection}
\end{figure}

\begin{figure}
\centering
\includegraphics[scale=0.4]{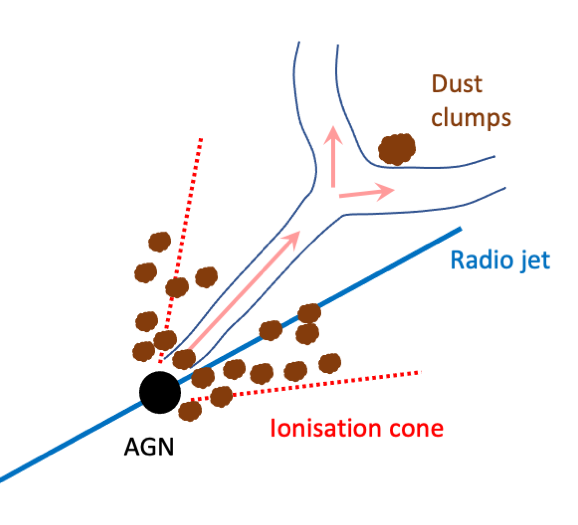}
\caption{A cartoon model of the ionised outflow observed with the MUSE-NFM data in the Circinus galaxy. The outflow might be launched as a collimated structure by the radio jet, which then fragments into two filaments probably due to the presence of a dense clump at the tip of the collimated structure. The dust is distributed along the ionisation cone, apparent from the extinction maps and also consistent with archival mid-infrared observations.}
\label{fig:final_picture}
\end{figure}

The results presented in Section \ref{sect4} highlights the complex structures within an outflow that are revealed from high resolution observations in the vicinity of the AGN. The presence of an outflow in the form of an ionisation cone in the Circinus galaxy has been known for nearly three decades, and thanks to the current state-of-the-art instrumentation that we are now able to resolve parsec scale emission using optical spectra. The origin of the initial clumpy collimated structure observed in the NFM data could be due to a small-scale radio jet. The Circinus galaxy is known to host a radio jet, although its PA is aligned closer to the edge of the ionisation cone \citep[e.g.,][]{elmouttie98}. However, these are lower spatial resolution radio imaging observations and the regions close to the AGN torus is not resolved. Due to the precession and interaction with the surrounding medium, jets in AGN are known to bend and change directons at larger scales \citep[e.g., as in the case of NGC 1068][]{gallimore04}. Therefore, future work will target the regions closer to the AGN to search for the potential impact of small-scale jets that may be aligned along the axis of the ionisation cone. The relative orientations of the radio jet and the ionisation cone in the Circinus galaxy are depicted in Fig. \ref{fig:final_picture}.

The outflow itself is not composed of uniformly distributed ionised gas, but shows the presence of clumps, confirming previous observations from lower resolution data of other targets that the outflowing media are non-uniform in nature \citep[e.g.,][]{kakkad18, kakkad22}. The presence of clumps or knots along outflow/ionised gas filaments in the Circinus galaxy has also been previously reported in \citet{veilleux97} and these structures have been attributed to bow-shocked features that resemble Herbig Haro objects that interact strongly with the surrounding ISM. The observed morphology in the NFM data could be a smaller-scale version of the observations in larger scales.

The collimated component of the outflow most likely fragments into two components because of the presence of an obstruction in the path of the outflow. In the NFM data, there is an indication of the presence of a dust clump via extinction maps obtained from the outflowing component of H$\alpha$ and H$\beta$ emission. Infrared observations targeting primarily the dust emission could confirm this scenario. Filament structures are also observed in images that trace gas across larger scales \citep[e.g.,][]{marconi94, veilleux97} and it is probable that the NFM data presented in this paper reveals the origin of the kiloparsec-scale filaments. Fig. \ref{fig:NFM_WFM_connection} shows the ionised gas morphology traced by the archival MUSE-WFM observations (background) and the emission from the outflowing \oiii ~component from the MUSE-NFM data (red contours). The spatial distribution of the filaments observed in WFM is strongly suggestive that their origin is to be found in the fragmented arms of the tuning fork observed in the NFM observations.

Fig. \ref{fig:final_picture} shows an overall cartoon model of the ionised outflow in the Circinus galaxy, that shows the ionised outflow that fragments into two filaments, probably due to the presence of a dense clump at the tip of the collimated structure. The dust itself is distributed around the ionisation cone and it envelops the lower velocity ionised gas within the conical structure. We speculate that the outflow itself might be launched due to the radio jet, which cannot be robustly confirmed based on the available archival radio observations, as mentioned earlier.

As the NFM observations have only recently targeted the nearby galaxies, it is unclear if such outflow structures and fragmentations within outflows are common. If this is indeed observed for majority of the galaxies, conventional outflow models that use ionisation cone morphology may need to be revised to account for the results from these high spatial resolution data. 

\section{Summary \& Conclusions} \label{sect6}

We presented the MUSE NFM observations of the Circinus galaxy at a spatial resolution of 0.1\arcsec ~($\sim$2 pc) that resolves the regions close to the AGN torus. We derived the properties of the ionised gas outflow using the \oiii$\lambda$5007 emission line and the dust distribution using Balmer Decrement. We follow a non-parametric approach to analyse the emission lines, so the derived properties are independent of the fitting functions used. The main results of this work is summarized below:

\begin{itemize}
\item The flux distribution of the systemic component of \oiii ~emission, defined by the velocity components within $\pm$300 km s$^{-1}$, shows a conical morphology, which is also observed in larger spatial scales up to hundreds of parsecs in the literature. Archival radio observations show that the radio jet is aligned approximately with the edge of the ionisation cone.  The flux distribution of the outflowing component of the \oiii ~emission, on the other hand, shows a collimated structure up to $\sim$30 pc before fragmenting into two arms that overall mimics a ``tuning-fork'' shape. The outflowing structure itself is not smooth and shows clumps at several locations. 
\item A comparison between the stellar kinematic map and the \oiii ~centroid map suggests that the ionised gas (both the systemic as well as the outflowing component) co-rotates with the host galaxy. Both the non-parametric velocity distributions, $v_{10}$ and $w_{80}$ show similar structures as the outflow flux i.e., the tuning-fork shapes. Most of this outflow is blue-shifted, consistent with the outflow models of the Circinus galaxy reported in the literature. 
\item We find a total instantaneous mass outflow rate of $\sim$0.01 M$_{\odot}$ yr$^{-1}$ (3$\times$10$^{-7}$ on average per pixel of size 0.5 pc) and a time-average mass outflow rate of 10$^{-4}$ M$_{\odot}$ yr$^{-1}$. This is much lower than the star formation within the galaxy and therefore, the ionised gas outflow is not expected to regular star formation within a radius of $\sim$100 pc from the AGN location. 
\item The extinction maps using the systemic components of the H$\beta$ and H$\alpha$ line show that the dust distribution is concentrated along the ionisation cone i.e., the polar direction, consistent with the archival mid-infrared observations. The extinction map obtained from the outflowing components of H$\beta$ and H$\alpha$ shows a scarce distribution, with a clump approximately at the tip of the collimated part of the outflow. This dust clump might explain the fragmentation in the outflowing ionised gas. 
\item We combine previous reported results gathered from the literature and the observed outflows from the NFM data presented in this paper to present a model of the ionised gas outflow in the Circinus galaxy. We suggest that the observed outflow in the Circinus galaxy is composed of high velocity collimated gas that is enveloped by lower velocity dusty ionised gas. The presence of a dust clump at the tip of the collimated part of the outflow might be responsible for the fragmentation in the outflowing gas. The collimated outflow might be launched by a radio jet. Although the jet, observed in low spatial resolution radio observations, does not show a 1:1 alignment with the outflowing cone PA, they are known to bend and change at larger scales.
\end{itemize}

While the MUSE-NFM only provides the kinematic information in the ionised gas phase, outflows have been known in exist also in the molecular gas phase in the Circinus galaxy. A multi-wavelength approach to trace gas in the other phases such as warm and cold molecular, at the same spatial resolution of $\sim$2 pc, will be key to obtaining a holistic view into the outflow-AGN connection in the Circinus galaxy. Furthermore, high spatial resolution radio observations will verify if the observed collimated outflow is a result of jet-ISM interaction on small scales. 

\section*{Acknowledgements}
We would like to thank the anonymous referee for insightful comments and suggestions to improve this manuscript. The authors also thank T. Fischer for useful discussions. Based on observations from the ESO programme ID 0103.B-0396. M.S. and S.K. acknowledge support by the Science Fund of the Republic of Serbia, PROMIS 6060916, BOWIE and by the Ministry of Education, Science and Technological Development of the Republic of Serbia through contract No.~451-03-9/2022-14/200002. D.A. acknowledges funding through the European Union’s Horizon 2020 and Innovation programme under the Marie Sklodowska-Curie grant agreement no. 793499 (DUSTDEVILS).

%%%%%%%%%%%%%%%%%%%%%%%%%%%%%%%%%%%%%%%%%%%%%%%%%%
\section*{Data Availability}
The data presented in this paper is available with ESO archive under the programme ID: 0103.B-0396.

%%%%%%%%%%%%%%%%%%%% REFERENCES %%%%%%%%%%%%%%%%%%

% The best way to enter references is to use BibTeX:

\bibliographystyle{mnras}
\bibliography{reference} % if your bibtex file is called example.bib

% Alternatively you could enter them by hand, like this:
% This method is tedious and prone to error if you have lots of references
%\begin{thebibliography}{99}
%\bibitem[\protect\citeauthoryear{Author}{2012}]{Author2012}
%Author A.~N., 2013, Journal of Improbable Astronomy, 1, 1
%\bibitem[\protect\citeauthoryear{Others}{2013}]{Others2013}
%Others S., 2012, Journal of Interesting Stuff, 17, 198
%\end{thebibliography}

%%%%%%%%%%%%%%%%%%%%%%%%%%%%%%%%%%%%%%%%%%%%%%%%%%

%%%%%%%%%%%%%%%%% APPENDICES %%%%%%%%%%%%%%%%%%%%%

%\appendix

%\section{Appendix A}

%Appendix here

%%%%%%%%%%%%%%%%%%%%%%%%%%%%%%%%%%%%%%%%%%%%%%%%%%

% Don't change these lines
\bsp	% typesetting comment
\label{lastpage}
\end{document}